\documentclass[twocolumn]{aastex631}
\hypersetup{linkcolor=red,citecolor=blue,filecolor=cyan,urlcolor=magenta}

\usepackage{CJK}
\usepackage{amsmath} 
\usepackage{bm}      
\usepackage{lmodern}
\usepackage{amssymb}
\usepackage{physics} 
\usepackage{pgfplots}  

\shorttitle{Density Distribution of Plasmas In Galaxies}
\shortauthors{Chen Haibin}

\graphicspath{{./}{figures/}}

\begin{document}
\begin{CJK*}{UTF8}{gbsn}
\title{Density Distribution of Plasmas Resembling Dark Matter Halo Due to Ionization lag and Ambipolar Electric Field}

\author[0000-0002-5500-3634]{Haibin Chen}
\correspondingauthor{Haibin Chen}
\email{chenhb3@mail2.sysu.edu.cn}

\author[0000-0003-0264-4363]{Rong Wu}

\begin{abstract}

%

In a spherically symmetric plasma constrained by its own gravity, the ionization degree lags behind changes in temperature and density. The ambipolar electric field accelerates ions radially and cools electrons. Ions lose energy and angular momentum in collisions with low-temperature electrons.  The angular momentum of ions decreases much faster than their energy in cycles. The trajectories of ions are close to radial, and the density distribution resembles pseudo-isothermal.

The velocity distributions of baryons moving outward and inward in galaxies tend to approach independent Maxwell distributions. This characteristic suppresses small-angle scattering of ions, reducing the ion collision cross-section by three orders of magnitude. In the massive galaxies, baryons can replace all dark matter.

\end{abstract}



\section{Introduction}

\citet{1933AcHPh...6..110Z} discovered that the velocity dispersion in the Coma Cluster was much higher than expected, suggesting the existence of a large amount of invisible mass, which was considered as preliminary evidence for dark matter. \citet{1970ApJ...159..379R} further confirmed this discovery through studies of the rotation curve of the Andromeda Galaxy, indicating that galaxies may be embedded in massive halos. 

The cold dark-matter model successfully explains both the emergence and evolution of cosmic structures on large scales  \citep{1985ApJ...292..371D}  \citep{1997ApJ...490..493N}  \citep{1996ApJ...462..563N}  \citep{2011ApJS..192...18K} \citep{2005Natur.435..629S}  . However, it encounters some difficulties in describing structures at galactic scales   \citep{1994Natur.370..629M}  \citep{1999MNRAS.310.1147M} \citep{2011MNRAS.415L..40B} \citep{2017ARA&A..55..343B} \citep{2017Galax...5...17D} \citep{2020Univ....6..107D} .

Referring to the motion state of cold dark matter, if baryons experience weak collisions during their motion and their trajectories are close to radial, their density distribution can also resemble that of a dark matter halo. This raises two questions: What dissipation mechanism can cause the density distribution of baryons to significantly deviate from equilibrium? Can this kinematic characteristic significantly reduce the Coulomb collision cross-section of ions?

\section{Ionization Hysteresis}

The ionization degree of plasmas needs to be altered through ionization or recombination reactions, and its changes lag behind temperature and density variations. This results in a mismatch between the ionized and recombination regions, leading to ambipolar diffusion and ambipolar electric fields.

According to the Saha equation\citep{chen1984introduction}, the ionization equilibrium equation for hydrogen is
\begin{equation}
\frac{n_i\cdot n_e}{n_H} =2.4\times 10^{15} T_e^{\frac{3}{2}} e^{-\frac{\varepsilon_H}{kT_e }}
\end{equation}
where $n_i$ is the number density of hydrogen ions, $n_e$ is the number density of electrons, $n_i=n_e$ when the plasma contains only hydrogen, $n_H$ is the number density of neutral hydrogen, $T_e$ is the electron temperature, and $\varepsilon_H=13.6ev$ is the ionization energy of hydrogen.

The ionization degree is denoted as $\beta =\frac{n_e}{n_H+n_e }$. It can be seen from the Saha equation that the ionization degree at equilibrium $\beta _0$ primarily depends on the temperature $T_e$ and is influenced by particle number density.

When a plasma enters a new region with different ionization degrees, such as moving radially outward, its temperature and density change, but it still maintains its original ionization degree. The ionization degree deviates from the equilibrium state, and the ionization rate $R_{io}$ and the recombination rate $R_{re}$ are not equal. As the reaction proceeds, the ionization degree changes, and the change in electron number density is
\begin{equation}
\frac{dn_e}{dt}=R_{io}-R_{re}
\end{equation}
The changes in the ionization degree $\beta$ of the plasma and the ionization lag $\delta \beta =\beta -\beta _0$ are
\begin{equation}
\frac{d\delta \beta }{dt}=\frac{d\beta }{dt}=\frac{R_{io}-R_{re}}{n_H+n_e }
\end{equation}
At ionization equilibrium, $R_{io}=R_{re}$. Near the equilibrium state, $R_{io}$ can also be expressed using $R_{re}$ and the ionization degree at equilibrium $\beta _0$. The above equation can be rewritten as
\begin{equation}
\frac{d\beta }{dt}=-R_{re} \delta \beta \frac{n_{H0}+n_{e0}}{n_{H0} n_e }=-R_{re} \delta \beta \frac{1}{\beta (1-\beta _0 )(n_H+n_e )}
\end{equation}
It can be seen that the change in plasma ionization degree depends on the plasma ionization and recombination rates. After changes in temperature and density, the ionization degree $\beta$ gradually approaches the ionization degree at equilibrium $\beta _0$ as the reaction proceeds.

The relationship between the velocity of the plasma diffusing outward from the central region of the galaxy and the rate of change of ionization degree is
\begin{equation}
\frac{d\beta }{dt}=\frac{\partial \beta }{\partial r} v_r\approx \frac{\partial \beta _0}{\partial r} v_r
\end{equation}
Rearranging this gives
\begin{equation}
\delta \beta =-\frac{v_r \beta (1-\beta _0 )(n_H+n_e )}{R_{re}}   \frac{\partial \beta _0}{\partial r}
\end{equation}
Substituting $R_{re}$ \citep{bell1981capture} and specific model data allows the calculation of $\delta \beta$. This formula is accurate under the condition $\delta \beta \ll 1$.

It can be seen that if the ionization degree gradient at plasma equilibrium $\frac{\partial \beta _0}{\partial r}<0$, when the plasma moves outward, its ionization degree is higher than the local ionization degree, and when the plasma moves inward, its ionization degree is lower than the local ionization degree. This is the phenomenon of ionization lag. If there is a radial electric field, electrons and ions in the plasma perform work during this process.

The ionization lag phenomenon leads to the separation of ionization and recombination regions in the plasma. Some neutral atoms move from low ionization regions to high ionization regions, ionize, and then return to low ionization regions as ions and electrons, recombining into neutral atoms in the low ionization regions, forming a cycle. This process can also be seen as ambipolar diffusion from high ionization regions to low ionization regions.

\section{Ambipolar Diffusion and Ambipolar Electric Fields in Galaxies}

Ionization lag leads to ambipolar diffusion, and stable ambipolar diffusion can generate a stable ambipolar electric field. Ambipolar diffusion cools electrons.

In plasmas, electrons have higher thermal motion velocities and a stronger diffusion trend. The electron density tends to decrease and form an electric field, which constrains the motion of electrons and maintains the electrical neutrality of the plasma. This spontaneous electric field generation and self-regulating diffusion behavior in plasmas is called ambipolar diffusion, and the electric field generated due to the different thermal motion velocities of electrons and ions is called the ambipolar electric field. \citep{chen1984introduction}

In galaxies, the particle density is extremely low, and collisions between electrons and neutral particles can be neglected. Ionization lag provides a stable plasma source, and the ambipolar electric field in galaxies can be calculated from the electron pressure gradient:
\begin{equation}
\vec{E}=\frac{\nabla p_e}{en_e}
\end{equation}
where $n_e$ is the electron density, and $p_e$ is the pressure exerted by the electrons:
\begin{equation}
p_e=n_e kT_e
\end{equation}
where $T_e$ is the electron temperature.

The presence of the ambipolar electric field causes electrons to perform work externally during expansion, and ionization retardation provides a stable electron source. If the density change of electrons due to expansion during ambipolar diffusion is very rapid, the power of expansion work is much greater than the power of recombination and ionization reactions (absorbing or releasing heat), and the relationship between electron temperature and density approximately satisfies an adiabatic relation:
\begin{equation}
T_e\propto n_e^{\frac{2}{3}}
\end{equation}
Obviously, in regions with higher electron density, the electron temperature is also higher. The relationship between electric potential and temperature satisfies:
\begin{equation}
U=\frac{5}{2} \frac{kT_e}{e}
\end{equation}
Electrons perform negative work in the electric field, transferring energy to ions. Therefore, in regions affected by ambipolar diffusion, the energy of electron is lower than the ion.

In galaxies, the density of the central region is greater than that of the outer region, and the electron temperature in the central region of the galaxy is higher. According to the Saha equation, the ionization degree in equilibrium is mainly determined by temperature. The ionization degree is higher in the central region of the galaxy in equilibrium and lower in the outer region. There is stable ambipolar diffusion from the central region to the outer region of the galaxy, and there is also an electric field along the radial direction of the galaxy. The initial model where electrons are in isothermal distribution will also become close to adiabatic distribution after ambipolar diffusion.

In real galaxies, ionization or recombination processes during ambipolar diffusion affect the electron temperature, so the relationship between electron temperature and density does not strictly satisfy the adiabatic relation, but the adiabatic model is more accurate than the isothermal model qualitatively.

%
%
%
%
%
%
%

\section{Energy Transfer Between Ions and Electrons}  
\label{sec:energy}

The most significant effect of ambipolar diffusion from the interior to the exterior of a galaxy is the cooling of electrons and the radial acceleration of ions. The energy transferred from electrons to ions (baryons) can accumulate, leading to an increasing energy difference between ions and electrons.

After the mutual transformation between ions and neutral atoms, the charge carried by them changes. They can be considered as baryons carrying different charges and subject to different electric field forces. Under the condition that the baryon's mean free path exceeds one motion cycle, its trajectory can be described using kinematics, and the effect of radial acceleration of ions can accumulate. If the mean free path is short, much shorter than one motion cycle, its motion can only be studied statistically, and the effect of radial acceleration of ions cannot accumulate.

Assuming that the baryon dark matter halo is spherically symmetric, due to ionization lag, the main ionization region with a radius of $r_{io} \pm \sigma r_{io}$ and the main recombination region with a radius of $r_{re} \pm \sigma r_{re}$ do not overlap, and $r_{io} < r_{re}$. There is ambipolar diffusion and an ambipolar electric field from the ionization region to the recombination region, directed radially outward. The average potential difference between the two regions is $U_{io} - U_{re}$. Electrons lose an average energy of $e(U_{io} - U_{re})$ during ambipolar diffusion, while protons gain energy $e(U_{io} - U_{re})$. The electric field mediates energy transfer between them. In this process, baryons are radially accelerated by the electric field, and the trajectories of radially accelerated baryons shift radially.

The typical temperature for plasma ionization in galaxies corresponds to the ionization energy of hydrogen, which is $\epsilon_H = 13.6 \, \text{eV}$. The typical temperature for recombination is related to ionization lag. In one cycle, the energy gained by ions, $e(U_{io} - U_{re})$, can be estimated using $\epsilon_H \delta \beta$, and we believe its magnitude is approximately $0.01$ to $10 \, \text{eV}$.

Under the condition that the baryon's mean free path exceeds one cycle, its trajectory can be predicted. The energy gained by baryons from ionization lag and the ambipolar electric field can accumulate. The temperature of electrons will decrease, and the energy of baryons will occupy most of the system's energy.

%
%
%
%
%
%
%

\section{Energy and Angular Momentum Balance of Baryons}

	\begin{figure}[htbp]  
	\centering  
	\includegraphics[width=0.5\textwidth]{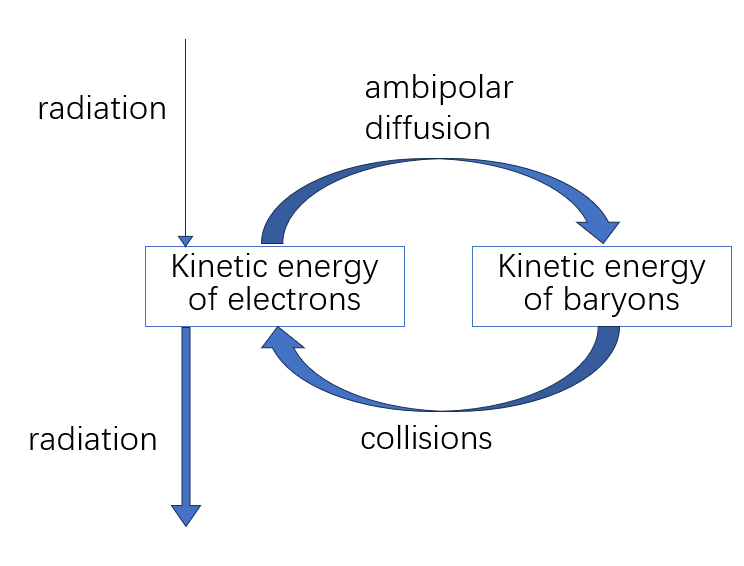}  
	\caption{energy cycle}  
	\label{energy}  
	\end{figure}

Ionization lag and ambipolar diffusion lead to a decrease in the temperature of electrons. Coulomb collisions between high-speed ions and low-temperature electrons have two effects: one is the reduction of ion velocity and heating of electrons, and the other is the change in the direction of ion motion, which belongs to small-angle scattering in Coulomb collisions. Both effects can alter the angular momentum of baryons. The relationship between the angular momentum and energy of baryons influences the mass distribution of baryons in galaxies.

In ambipolar diffusion, energy is transferred from electrons to ions, while in Coulomb collisions, energy is transferred from ions to low-temperature electrons. Therefore, an energy cycle is established between ions and electrons through collisions and ambipolar diffusion. The flow of energy is illustrated in the figure\ref{energy}. Electrons lose energy through radiation,, and electrons can also absorb radiation to increase their energy, but the absorbed energy is much lower than the lost energy. Overall, the mechanical energy of the system is continuously decreasing. In the energy cycle between baryons and electrons, there is a new balance of angular momentum for baryons.

After Coulomb collisions between ions and low-temperature electrons, ions are decelerated, and both their angular momentum and energy are reduced and transferred to electrons. After electrons gain energy, some of it is lost through radiation, while some is returned to ions through ionization lag and the ambipolar electric field. Therefore, the rate of energy reduction for ions is relatively slow, comparable to the rate of mechanical energy reduction for the entire system. After electrons gain angular momentum, it is quickly averaged out through collisions with other electrons. Since ion energy can be returned but angular momentum is difficult to return, the rate of angular momentum reduction for ions is much faster than the rate of energy reduction.

Due to the significantly shorter mean free path of low-temperature electrons compared to high-speed ions, the motion of electrons is irregular, and the angular momentum distribution of electrons is determined by temperature and position. However, the mean free path of high-speed ions can be very long, and the angular momentum distribution of ions can be shaped by the motion process.

As angular momentum decreases, the motion direction of ions in most regions becomes nearly parallel to the radius. The change in angular momentum accompanying the deceleration of ions becomes smaller and smaller until a new balance is reached. If small-angle scattering in Coulomb collisions of ions is suppressed, the angular momentum of ions is comparable to that of electrons, depending on the angular momentum of electrons along the entire motion path. If small-angle scattering is not negligible, collisions between ions and low-temperature electrons will cause random changes in the direction of motion and angular momentum of ions, making it difficult for ions to aim at the center of the galaxy. Although the variance of ion angular momentum is low, it will not decrease to the level of electrons.

%
%
%
%
%
%
%
%
%
%
%
%

\section{Baryon's Trajectories and Density Distribution}  
	\begin{figure}[htbp]  
	\centering  
	\includegraphics[width=0.5\textwidth]{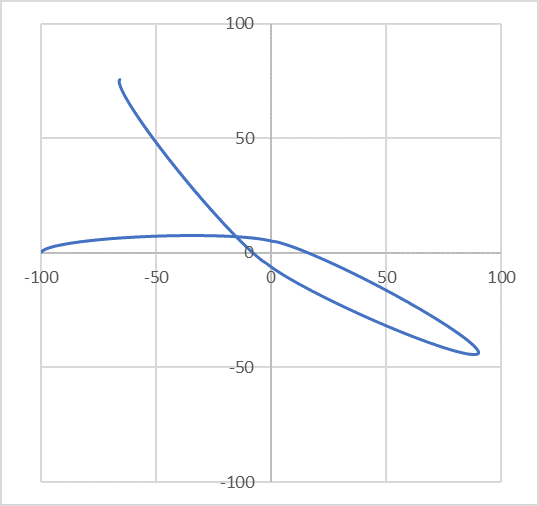}  
	\caption{trajectory of a baryon}  
	\label{trajectory}  
	\end{figure}

The closest distance $r_{\min}$ and the farthest distance $r_{\max}$ of a baryon's trajectory to the galactic center are two intuitive parameters influenced by the key parameters of energy $E$ and angular momentum $J$. The trajectory of a baryon can affect its mass distribution in the galaxy.

Under the influence of the galactic electric field and ionization lag, the energy of baryons decreases relatively slowly, while the angular momentum decreases rapidly. This is reflected in the trajectory, where $r_{\max}$ of the baryon's trajectory decreases slowly, while $r_{\min}$ decreases rapidly to nearly 0, causing $r_{\min}/r_{\max} \rightarrow 0$.

In gravitational systems with mass concentrated at the center, such as the solar system, celestial bodies with $r_{\max}/r_{\min} \gg 1$ have trajectories with eccentricities close to 1, like the trajectory of a comet. However, in galaxies, where mass is not concentrated at the center, the trajectories of celestial bodies are not conic sections. The bending of a baryon's trajectory near $r_{\min}$ is not 180°, and the turning angle is generally small. A typical trajectory of a baryon is shown in the figure \ref{trajectory}.

In the region where $r_{\min} < r < r_{\max}$, the mass distribution of a galaxy composed of such baryons is close to:
\begin{equation}
\rho(r) \sim \frac{1}{v_r r^2}
\end{equation}
where $v_r$ is the radial velocity of the baryon.

A galaxy composed of baryons with $r_{\max} \approx R_0$ and $r_{\min} \approx 0$ is a good simplified model. Using $v_r = \text{constant}$ as the initial input, within the range $(0, R_0)$, the density distribution is:
\begin{equation}
\rho(r) \sim \frac{1}{r^2}
\end{equation}
Based on this density distribution, $v_r(r)$ can be solved. After multiple iterations of radial velocity $v_r(r)$ and density $\rho(r)$, a more accurate density distribution of this model can be obtained, and higher-order iterations can be completed through numerical calculations.

\begin{figure}[htbp]  
	\centering  
	\includegraphics[width=0.5\textwidth]{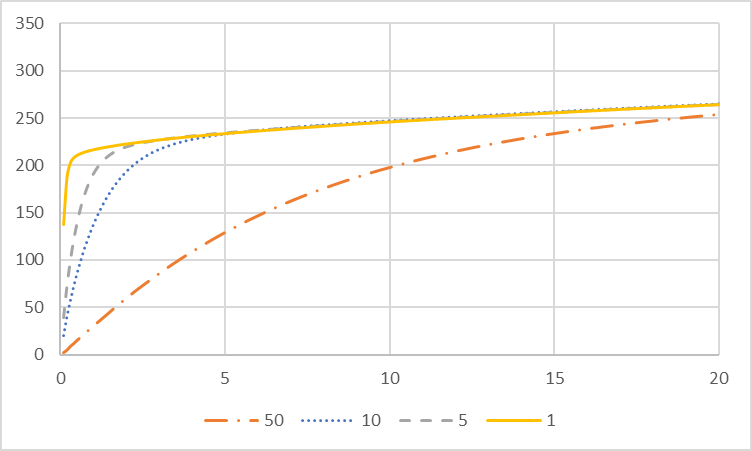}  
	\caption{rotation curves of galaxies}  
	\label{rotation}  
	\end{figure}

The closest distance $r_{\min}$ of a baryon's trajectory to the galactic center is not zero but has some scatter. We can construct similar galaxies composed of baryons with $r_{\max} \approx R_0$ and $r_{\min}$ distributed according to a two-dimensional normal distribution (two-dimensional Maxwell distribution). Following the previous iteration process, the galaxy rotation velocity $v_c$ curve obtained through numerical simulation is shown in the figure. The trajectory figure \ref{rotation} is also obtained through this numerical simulation. The code for the simulation program is included in the source file `fortranprogram.txt`.

In the simulation, $R_0 = 100$, and only a segment from 0 to 20 is shown in the figure. Different curves in the figure are labeled with the dimensionless parameter $\frac{v_0 \sqrt{R_0}}{\sqrt{GM}}$, where $v_0$ is the characteristic velocity perpendicular to the radial direction at $r_{\max}$, and $M$ is the total mass of the galaxy. Selecting the region near the center as the research object, we can obtain a nearly flat rotation curve.\citep{2008AJ....136.2648D} The simulation results are close to the pseudo-isothermal model.

\begin{figure}[htbp]  
	\centering  
	\includegraphics[width=0.5\textwidth]{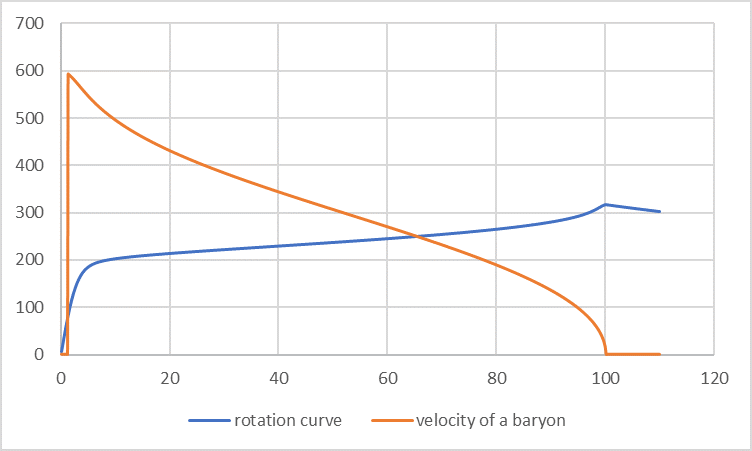}  
	\caption{velocity of a baryon and rotation curve}  
	\label{velocity}  
	\end{figure}

Baryons reaching near the galactic center from the outskirts of the galaxy are accelerated by gravity. In the simulation, we found differences in the distribution of baryon dark matter velocity $v_{bd}$ and rotation velocity $v_c$, as shown in the figure \ref{velocity}. This is a comparison of baryon velocity and the rotation curve. (A baryon velocity of 0 indicates that the baryon cannot reach that region.)

In spiral galaxies and barred spiral galaxies, the rotation velocity $v_c$ is mainly measured through HI linewidth measurements. The velocity of baryon dark matter $v_{bd}$ may appear in the HI radiation background. In elliptical galaxies, the measured result may be $v_{bd}$ rather than the rotation velocity $v_c$. Near the galactic core, the value of $v_{bd}$ is generally 1 to 4 times the value of the flat segment of the rotation curve. This may lead to an overestimation of the mass and mass-to-light ratio of elliptical galaxies.

In summary, the galactic electric field and ionization lag have significant effects on the energy and angular momentum of baryons and electrons, ultimately making the mass distribution of baryons similar to that of the dark matter halo. In the case of only baryons, a model similar to the observed galactic rotation curve can also be obtained. Baryons can replace some or all of the dark matter in galaxies.

%
%
%
%
%
%
%
%
%
%
%
%
%

\section{Coulomb collisions and mean free path of ions  }

In the context of galaxy astronomy and plasma physics, when collisions are strong, the trajectories of baryons are irregular. It is only under weaker collisions that the effects of ionization relaxation and ambipolar diffusion on the radial acceleration of baryons can be preserved. In the course of research, to ensure regular trajectories of baryons, we introduce the assumption of weak collisions. This assumption becomes a crucial constraint when attempting to partially or fully replace the mass of dark matter halos with baryonic dark matter.  

One primary pathway for baryon collisions is Coulomb collisions between ions and other charged particles. When in neutral hydrogen form, collisions with electrons, ions, and neutral hydrogen mainly lead to ionization, with negligible impact on its trajectory.  

The cross-section for a single collision between an ion and another charged particle resulting in a deflection greater than $\pi/2$ is given by \citep{krall1973principles}
\begin{equation}  
\sigma_n=\frac{1}{16\pi} \left(\frac{e^2}{\epsilon_0 m_i v_i^2}\right)^2=C_1 \frac{1}{v_i^4}  
\end{equation}  
This cross-section is known as the large-angle scattering cross-section, where $\epsilon_0=8.854\times10^{-12} \text{F}\text{m}^{-1}$, or $\text{C}^2 \text{m}^{-3} \text{kg}^{-1} \text{s}^2$ is the vacuum permittivity, $m_i=1.673\times10^{-27} \text{kg}$ is the proton mass, and $e=1.602\times10^{-19} \text{C}$ is the elementary charge. The coefficient $C_1$ is calculated to be $0.05972\text{m}^{-6} \text{s}^4$.  

Due to the long-range nature of Coulomb interactions, small-angle collisions are generally more frequent. The cumulative effect of many small-angle deflections is greater than that of large-angle collisions. The collision cross-section is  
\begin{equation}  
\sigma_f=\frac{1}{2\pi} \left(\frac{e^2}{\epsilon_0 m_i v_i^2}\right)^2 \ln\Lambda=8\ln\Lambda\sigma_n=C_2 \frac{1}{v_i^4}  
\end{equation}  
where $\ln\Lambda$ is typically taken as 10, and $\Lambda$ is the ratio of the Debye radius to the near-collision radius. The cross-section for small-angle scattering is $8\ln\Lambda$ times higher than that for large-angle scattering, differing by about two orders of magnitude. $C_2=4.778\text{m}^{-6} \text{s}^4$.  

Assuming that the mass distribution of the dark halo of a galaxy can be described by a pseudo-isothermal model \citep{2003MNRAS.339..243J},  
\begin{equation}  
\rho_{\text{iso}}(r)=\frac{\rho_0}{1+(r/r_c)^2}  
\end{equation}  
then the rotation curve is  
\begin{equation}  
v_{\text{iso}}(r)=v_h \sqrt{1-\frac{r_c}{r} \arctan\left(\frac{r}{r_c}\right)}  
\end{equation}  
where  
\begin{equation}  
v_h^2=\rho_0 4\pi r_c^2 G  
\end{equation}  
If the proportion of baryonic dark matter in dark matter is $\eta$, the characteristic particle number density corresponding to the characteristic density $\eta\rho_0$ of baryonic dark matter in the central region of the galaxy is  
\begin{equation}  
n_0=\frac{\eta\rho_0}{m_H} =\frac{\eta}{4\pi m_H G} \frac{v_b^2}{r_c^2} =7.127\times10^{35} \eta \frac{v_h^2}{r_c^2}  
\end{equation}  
where the length unit in the equation is meters.  

The collision impact of baryons within the central region $r_c$ of the galaxy is relatively small, while collisions of baryons outside $r_c$ can significantly alter their angular momentum and change their trajectories. To satisfy the assumption of weak collisions, the collision probability of baryons outside $r_c$ needs to be low.  

Assuming that baryons move outward from $r_c$ and return to $r_c$ after nearly one revolution, and their velocity $v_i$ near $r_c$ is equal to $C_v v_h$, the expected number of collisions with electrons during this process is  
\begin{equation}  
nc=2r_c \eta n_0 \sigma_n=6.811\times10^{36}\times\frac{\eta}{r_c} \frac{1}{C_v^4 v_h^2}  
\end{equation}  
Since the relative velocity between ions and oppositely moving ions is double that between electrons and ions, collisions between ions moving in different directions are an order of magnitude lower and can be neglected for now.  

For larger galaxies, we can take $v_h=500\text{km/s}$ and $r_c=10\text{kpc}=3.086\times10^{20} \text{m}$. According to numerical simulation results, we temporarily take $C_v=3$, and the expected number of collisions is  
\begin{equation}  
nc=1090\eta  
\end{equation}  
If the weak collision assumption requires $nc<1$, then the proportion of baryonic dark matter $\eta$ is less than $9\times10^{-4}$. If one wants to replace all dark matter with baryons, the mean free path of ions needs to be increased by nearly three orders of magnitude.  

We believe that under the condition of satisfying the weak collision assumption, the collision characteristics of baryonic dark matter still allow room for increasing its proportion $\eta$.

\section{Failure of small-angle scattering among ions:}

Due to ionization lag and ambipolar electric fields, baryons experience radial acceleration, resulting in radial velocities significantly larger than velocities perpendicular to the radial direction. This is analogous to a model where two types of low-temperature ions collide at high speeds, one moving outward and the other inward. Assuming a relative velocity of $2v_r$ for the colliding plasmas moving outward and inward, and an average thermal velocity within the plasma of $v_T$, if $v_r \gg v_T$, their velocity distributions separately satisfy the Maxwell distribution, and small external perturbations only cause minor deviations from this distribution. Small-angle scattering in Coulomb collisions becomes ineffective.

The cross-section for Coulomb collisions between charged particles is proportional to $v^{-4}$, the collision frequency is proportional to $v^{-3}$, and the particle momentum exchange rate is proportional to $v^{-2}$, where $v$ is the relative velocity between the charged particles. The momentum exchange rate for an ion with a relative velocity of $v$ and a total number of ions $n$ is $nv^{-2}$. Clearly, the higher the relative velocity, the lower the momentum exchange rate. Only when $dn/dv \sim v^2$ do we need to consider momentum exchange with ions at higher velocities. If the velocity distributions of the two types of ions are isolated from each other, Coulomb collisions between them do not need to be considered.

For example, in an outward-moving ion cloud, if their thermal velocities are small and their outward velocities are large, all ions have radial velocities greater than 0, correspondingly, all inward radial velocities are less than 0. The velocity distributions of these two types of ions are isolated from each other. The momentum exchange rate between outward-moving ions and inward-moving particles is much smaller than their internal momentum exchange rates. Outward-moving ions tend spontaneously towards a Maxwell distribution, and the same is true for inward-moving ions. The velocity distributions of these two independent ions are stable. According to the $3\sigma$ principle of the normal distribution, when $v_r > 3v_T$, the velocity distributions of outward-moving ions and inward-moving particles can be considered independent.

In small-angle scattering in Coulomb collisions, momentum needs to accumulate continuously before the deflection angle exceeds $\pi/2$ and is counted as a collision. As a example, in a galaxy, if $v_r \gg v_T$, an outward-moving ion, after a weak Coulomb collision with an inward-moving ion, experiences a random change in its velocity perpendicular to the radial direction. If there are no other influencing factors, this outward-moving particle continues to undergo Coulomb collisions with inward-moving particles, with its velocity direction undergoing a random walk and gradually deviating from the radial direction until reaching a deflection of $\pi/2$. The deflection of the velocity direction is based on their relative motion of $2v_r$. When the outward-moving particle reaches a deflection of $\pi/2$, its radial velocity is $-v_r$, and its velocity perpendicular to the radial direction is $2v_r$.

Small-angle scattering is not accomplished in one instance. During the accumulation of deflection angles, the ion's velocity perpendicular to the diameter direction gradually increases, and then there is a trend to deviate from the original velocity group. When the ion's velocity is at the edge of this velocity group, collisions with ions in this group will push the ion's velocity towards the average velocity. Among these two effects, the effect that prevents ions from deviating from the original velocity group is stronger, so small-angle scattering becomes ineffective. Since the momentum exchanged in small-angle scattering is perpendicular to the relative velocity direction between ions, the failure of small-angle scattering involves both the failure of momentum exchange and energy dissipation.

Large-angle scattering in Coulomb collisions is accomplished in one instance, so it is not affected by isolated velocity distributions. After the failure of small-angle scattering, it becomes dominant in Coulomb collisions.

Collisions between electrons and ions also affect the motion of ions. The collision frequency between ions and electrons is greater than the collision frequency between ions and ions. In the direction perpendicular to the radius, the kinetic energies of ions and electrons are similar. In this direction, ions and electrons belong to the same velocity group, and collisions between electrons and ions also prevent ions from deviating from their original velocity group in small-angle scattering. In the radial direction, since the thermal velocity of electrons is generally greater than $v_r$, the velocity distribution of electrons belongs to the same velocity group as both outward-moving and inward-moving ions, making electrons a medium for the interaction between these two types of ions. Collisions with electrons reduce the relative velocity between these two types of ions and heat the electrons. This is the content studied in \ref{sec:energy} of this paper.

In the case where small-angle scattering among ions fails and only large-angle scattering between ions with different directions of motion is effective, the expected number of ion collisions is significantly reduced. We only need to calculate large-angle scattering between ions, and the formula for collision expectation becomes:

\begin{equation}
nc = 2.661 \times 10^{33} \times \frac{\eta}{r_c} \times \frac{1}{C_v^4 v_h^2}
\end{equation}

Here, due to the distinction in ion motion direction (outward or inward), the number density of ions moving in either direction is half of the total ion number density. The ratio of the collision cross-section for large-angle scattering to that for small-angle scattering is $1/8\ln\Lambda$. The relative velocity of particles moving in opposite directions is twice the particle velocity. The final calculated $nc$ is the expected number of Coulomb collisions considering only large-angle scattering between ions, and its coefficient differs from the expected number of Coulomb collisions between ions and electrons by approximately 2560 times.

For a large, massive galaxy with $v_h = 500 \text{ km/s}$, $r_c = 10 \text{ kpc}$, and $C_v = 3$, the expected number of collisions for baryons outside $r_c$ is:

\begin{equation}
nc = 0.4258 \eta
\end{equation}

When baryons replace all dark matter in a massive galaxy, the expected number of large-angle scattering collisions per cycle is $nc = 0.4258$. We believe this collision probability is basically satisfy the weak collision assumption. If $C_v = 4$ or a larger value is taken, a stricter weak collision assumption can be satisfied.

\section{Weak collision assumption for lower-mass galaxies}

If we take $C_v=4$, medium-mass galaxies with rotational velocity $v_h>250\text{km/s}$ and $r_c>5\text{kpc}$ can also replace all dark matter with baryons under the weak collision assumption. However, $C_v=4$ requires a larger ratio of the maximum to minimum distance of baryons, with the farthest baryons more concentrated in the outer regions.

Further increasing $C_v$ and $r_c$ can make lower-mass galaxies satisfy the weak collision assumption. However, the velocity of baryons in dwarf galaxies is too low, making it difficult to satisfy the weak collision assumption in a conventional way. One feasible method is to introduce exogenous baryonic dark matter into low-mass galaxies, formed by the focusing of dark matter halos from more massive galaxies under gravitational disturbances. Coincidentally, due to the baryons in the environment having velocities concentrated in one direction, the aggregated mass distribution naturally exhibits a bar structure. Another feasible method is to partially abandon the weak collision assumption.

Analysis reveals that after the failure of small-angle scattering of ions, as long as the electron temperature is low, baryonic dark matter can still maintain a state of high mechanical energy and low angular momentum when the weak collision assumption is not satisfied. However, the mass distribution in the central region of dwarf galaxies that do not satisfy the weak collision assumption is uniform, while the central region density of massive galaxies that satisfy the weak collision assumption can have a peak. This provides a possible explanation for the cusp-core problem.

Energy dissipation in galaxies that do not satisfy the weak collision assumption may be relatively high, leading to insufficient energy transfer from electrons to ions and requiring supplementary cooling mechanisms such as electron radiation. A surprising result is that if the speed of electron radiative cooling is sufficiently fast, a baryonic distribution similar to a dark matter halo can be obtained without ionization lag and galactic electric fields.

%
%
%
%
%

\section{Discussion}

There are severe conflicts between the parameters of the model presented in this paper and some measurement results. However, due to changes in the underlying mechanisms, such conflicts can be explained from the perspective of the new model. The new model also provides a direction for solving other unresolved issues.

1. Measurement of Electron Temperature

In this model, the energy of ions is much higher than that of electrons. In some regions, the speed of ions is comparable to that of electrons, which may lead to errors in indirectly measuring electron temperature through collisional excitation by heavy ions.

2. The Bullet Cluster

The Bullet Cluster (1E 0657-558),\citep{2018PhR...730....1T} with its separation of mass center and radiation region, is considered evidence of the existence of dark matter. The dark matter collision cross-section calculated based on the Bullet Cluster rules out baryons as the main component of dark matter.\citep{2004ApJ...606..819M} We believe it is necessary to consider the impact of the failure of small-angle scattering during galaxy collisions and recalculate the collision cross-section of dark matter.

3. Baryon Acoustic Oscillations

In baryon acoustic oscillations,\citep{2015PhRvD..92l3516A} ionization lag can convert the thermal energy of electrons into the kinetic energy of baryons, enhancing baryon acoustic oscillations, which may have a significant impact on the formation of large-scale structures in the universe.

4. Galactic Spiral Arms

In the galactic disk, ionization lag and ambipolar electric fields can convert the thermal energy of electrons into the kinetic energy of baryons, which is a two-dimensional version of the baryonic dark matter model for galaxies. It can enhance and drive the motion of baryon density waves, cool electrons, promote star formation, and is an important mechanism for the formation and motion of galactic spiral arms.\citep{1969ApJ...155..721L}

5. Active Galactic Nuclei

Baryons falling into black holes can produce intense electromagnetic waves, heating the electrons around the black hole. Under certain conditions, this mechanism can replenish the energy lost due to plasma recombination. The trajectories of baryons are heavily influenced by collisions, which may result in a higher accretion rate for black holes compared to other dark matter models. It is one of the possible mechanisms for the formation of quasars. \citep{1993ARA&A..31..473A}

6. Coronal Anomalous Heating

Large-scale high-speed baryons bombarding the sun can increase the temperature of the solar corona, which is a possible mechanism for coronal anomalous heating.\citep{2015RSPTA.37340259T} The transfer of electron energy to ions through ionization lag and ambipolar diffusion in the plasma of the solar corona is also a possible mechanism.

7. Radially decelerated ions

It is noteworthy that when the ionization gradient increases with radius, ions are decelerated along the radial direction, while electrons are heated, which is an anomalous situation.

8. Matter Acquisition

In the era of interstellar colonization, we can capture baryonic dark matter and use it as the most important source of matter and energy. This is a method to acquire a large proportion of the mass of this galaxy.

%
%
%
%
%
%
%
%
%
%
%
%
%
%
%
%
%
%
%
%
%
%
%

\section{Conclusion}

The ionization degree of plasma in galaxies is influenced by temperature and density, with an ionization gradient existing along the radial direction. The ionization degree of plasma moving along the radial direction lags behind changes in temperature and density. The inconsistency between ionization and recombination regions leads to ambipolar diffusion from high ionization regions to low ionization regions in galaxies. The relationship between the temperature of moving electrons and density is close to adiabatic. The ionization degree of galactic plasma decreases with increasing radius. Ambipolar diffusion outward along the radial direction cools electrons under the effect of the ambipolar electric field and accelerates ions along the radial direction. In collisions with low-temperature electrons, ions transfer energy and angular momentum to electrons. The decrease in angular momentum of ions is much faster than that of energy. This makes the trajectories of baryons close to the radial direction, forming a density distribution similar to a pseudo-isothermal model.

The model requires a large free path for neutral hydrogen or ions, at least comparable to one cycle. In massive galaxies, the proportion of plasma as dark matter can only be less than $9\times10^{-4}$. However, the thermal velocity of this plasma is significantly smaller than the radial velocity. The smaller the relative velocity between charged particles, the more intense the collisions, which makes ions moving inward and outward spontaneously tend to independent Maxwell distributions, respectively. The collision rate and angular momentum exchange between ions moving in different directions are much smaller than those between ions moving in the same direction. The cumulative process of small-angle scattering is interrupted, and only large-angle scattering, which is a one-time process, is effective. The dark matter in massive and medium-sized galaxies can be entirely replaced by baryons. In dwarf galaxies, the free path of ions is shorter, and we have only discussed the direction for solving this problem. Finally, we discuss issues related to dark matter, such as the Bullet Cluster.

%
%
%
%

\section*{acknowledgments}
The data underlying this article are available in the article and in its online supplementary material.

\bibliography{reference}{}
\bibliographystyle{aasjournal}

\end{CJK*}
\end{document}